\documentclass[b5paper,twoside]{jpconf}  
\usepackage{graphicx}

\newcommand{\beq}{\begin{equation}}
\newcommand{\eeq}{\end{equation}}
\newcommand{\beqn}{\begin{eqnarray}}
\newcommand{\eeqn}{\end{eqnarray}}

\begin{document}
 
\title[Neptune's Migrations and Resonant Captures]
{Modeling the Migration of Neptune and the 
       Corresponding Resonant Captures}  

\author[Yeh, Jiang, Zhou]{Li-Chin Yeh$^1$, Ing-Guey Jiang$^2$,
 and Li-Yong Zhou$^{3}$}

\address{$^1$Department of Applied Mathematics,\\
National Hsinchu University of Education, Hsin-Chu, Taiwan
}
\address{$^2$Department of Physics and Institute of Astronomy,\\
National Tsing-Hua University, Hsin-Chu, Taiwan
}
\address{$^3$Department of Astronomy
 \& Key Laboratory of Modern Astronomy 
and Astrophysics in Ministry of Education,\\
Nanjing University, Nanjing 210093, China
}
\ead{jiang@phys.nthu.edu.tw} 

\begin{abstract}
Due to the angular momentum exchange with 
planetesimals, Neptune might have migrated outward to the
current position, and captured many Kuiper belt objects
(KBOs) into resonances. 
We set up a semi-analytic model to simulate the 
outward migration of Neptune, and the processes of resonant captures.
Our model can naturally explain Neptune's currently 
observed semi-major axis and eccentricity. The results 
show that the current population ratio 
between 3:2 and 2:1 is mainly due to the original density distribution
of KBOs, which might be related to drag-induced inward 
migrations of proto-KBOs.
\end{abstract}

\section{Introduction}

The possibility of planetary adiabatic
migrations due to the angular momentum exchange with planetesimals
in the Solar System
was proposed by Fernandez \& Ip (1984). As Jupiter is the
innermost gas giant planet and also the most massive one, it is
very effective to scatter planetesimals outward. Thus, the
Neptune, Uranus, and Saturn will have opportunities to gain
angular momentum from those scattered outward planetesimals.
Through the gravitational interaction with planetesimals, Jupiter
made inward migration and the others, i.e. Neptune, Uranus,
and Saturn made outward migrations (Fernandez \& Ip 1984).
In other words, Jupiter indirectly transferred angular
momentum to Neptune, Uranus, and Saturn through planetesimals.
However, as these planetesimals are much smaller than giant
planets, the amount of transferred angular momentum between
planetesimals and a planet is extremely small in each
scattering event, and this kind of migration must be a slow
adiabatic process.

In order to explain the orbital configuration of Pluto, Malhotra
(1993) proposed that Neptune has ever migrated outward and showed
that during this migration the mean motion resonances swept
through the planetesimal disc and capture Pluto into the 3:2
resonance, leading to an eccentric orbit.
Much more details of the above theory have been provided in
Malhotra (1995). For example, explicitly, an exponential orbital
migration was suggested as \beq a(t)=a_f - \Delta a\  {\rm
exp}(-t/\tau), \eeq where $a(t)$ is Neptune's semi-major axis at a
given time $t$, $a_f$ is Neptune's final semi-major axis, $\Delta
a$ is total change of semi-major axis during migration, and $\tau$
is the migration timescale. Through that assumption on Neptune's
orbital migration, Malhotra (1995) successfully showed that many
particles in the trans-Neptunian region could be captured into
resonances. For an orbital migration following Eq.(1), it will be
called as {\it Malhotra Migration} hereafter in this paper.

Note that Malhotra (1995) also described the heuristic picture
that how Neptune could gain angular momentum from planetesimals.
The details might be slightly different from what we just
mentioned above, but the principle is the same, i.e. Jupiter is
very effective in scattering planetesimals outward, and thus
Jupiter provides the angular momentum to Neptune indirectly
through planetesimals.

Zhou et al.(2002) added a stochastic term on the above exponential
orbital migration, and studied the effect of this term on the
outcome of resonant captures. They successfully showed that the
simulation results can be more consistent with the observational
ones, in the sense that less particles are captured into the 2:1
resonance than into the 3:2 resonance for some particular values
of the coefficient of this stochastic term. For this model of
Neptune's migration, it is called as {\it Zhou Migration}
hereafter in this paper.

However, the orbital eccentricity of Neptune is not mentioned in
the above two models. In both of the models, the migration
was attained by acting an artificial force on the planet and the
force was chosen in such a way that the orbital eccentricity of
planet is not affected. Thus Neptune holds current eccentricity
during the migration.
The orbital eccentricity shall be
an important parameter during the outward migration. In fact,
Neptune does have an eccentricity about 0.00858. It would be a good
question whether this eccentricity had affected the results
of resonant capture or not. It is even more interesting to study
whether this eccentricity can be reproduced or not during
Neptune's outward migration in a good simulation model.

In order to be more realistic, Gomes et al. (2004) employed about
10000 equal mass particles to represent planetesimals, which
interact with Neptune and push Neptune to do the outward
migration. They simulated and studied the details of Neptune's
outward migration for both high mass (about 100 to 200
$M_{\oplus}$) and low mass (from 40 to 50 $M_{\oplus}$)
planetesimal discs. The exponential behavior of outward migrations
is successfully produced in these simulations. Recently Li et al (2011) 
improve this model by including the self-gravitation of the planetesimals 
and the similar results are obtained. For a model
like this, it will be called as {\it Gomes Migration} hereafter in
this paper.

The main problem of Gomes Migration is that each planetesimal
particle's mass in their simulations are unrealistically larger
than the real ones. Moreover, the real planetesimals would not
have equal masses, but approximately have masses varying as a
power law. We shall note that due to the limitation of
computing capacity, it is difficult nowadays to reproduce
precisely the interactions between a planet with numerous
planetesimals with different sizes and masses, like what happens
in the real disc.

On the other hand,
in Zhou Migration model, they argued that the stochastic behavior 
in Neptune's migration had decreased the capture efficiency of 
the 2:1 resonance. Such a scenario was confirmed by Levison 
\& Morbidelli (2003). However, it is not exactly known whether 
this ``jumpy'' migration, as apparently observed in the numerical 
simulations by Gomes (2004) and Li et al. (2011), is a physically 
real behavior or it is just the artificial effects arising from the 
unrealistic large planetesimals applied in the simulations.  
In a theoretical model,
the parameter 
such as the maximum radius of the scattering planetesimal, 
depends on the details of the disc status in the early stage of 
the Solar System. We will leave these details to be addressed in 
Jiang et al. (2012). 

The above three migration models give very good implications on
the possible dynamical evolution of Neptune and other related
members in that region. Both Zhou Migration and Gomes Migration
are valuable methods to improve the theoretical modeling. In
order to find a physically meaningful way to model
close encounters without the trouble of massive n-body
numerical simulations of real particles, we here develop a
semi-analytic model to simulate Neptune's outward migration. The
driving mechanism is the angular
momentum exchange through encounters between Neptune and
planetesimals, which is the same as previous models. 
We use a semi-analytic method to treat
the encounters, so that the related physical parameters can be
meaningfully and reasonably controlled in an easy way. We will investigate
how the migration will be for different sizes or
distributions of planetesimals. Moreover, Neptune's initial
orbital eccentricity is set to be zero and it will evolve
naturally. Whether the final eccentricity is close to the
current observed value will be a criterion to determine
favored models. Further, the outcome of resonant captures through
outward migration of Neptune in our semi-analytic models will also
be investigated.

We would present our model in \S 2. 
The results and conclusions are in \S 3.

\section{The Model} 

The general picture is that Neptune moves inside a sea of planetesimals
at the outer Solar System. The encounters between Neptune and
planetesimals occur along any possible relative directions
randomly. Statistically, 
there are more scattered out-going
planetesimals than inward ones, so the net effect is
that Neptune gains angular momentum.
Thus, encounters in our model are assumed to
be along Neptune's orbit and to increase Neptune's
angular momentum.

We integrate Neptune's orbit numerically, subject to the Sun's
gravitational force and the kick force from planetesimals during
encounters. Planetesimals are not modeled dynamically. They are
considered as a reservoir of particles with fixed size and density
distributions. Through a Monte Carlo process, Neptune gets kicked
by planetesimals occasionally. The probability of encounters with
a particular kind of planetesimals will depend on the assumed size
distribution. Moreover, the encounter probability is certainly
smaller when Neptune moves to outer region where the number
density of planetesimals is smaller.

Many physical quantities of this planetesimal disc could influence
the overall outcome of random encounters. For example, the size
distribution of planetesimals,
the density distribution of the planetesimal disc, and
velocity distributions of planetesimals.

Naturally, the possible close encounters between Neptune and
planetesimals could occur along any relative directions
in a three-dimensional space. However, because
Jupiter scatters out many planetesimals effectively,
as a net effect, Neptune will
gain angular momentum from these out-going planetesimals.
This process is due to a statistical
average of the inward-outward
unbalance of planetesimals.
In order to simplify the model,
we assume that the net effect of close encounters
is dominated by the tangential close approach along Neptune's orbit.
In such an statistical unbalance,
the more out-going planetesimals than inward ones,
the more tangential close approach shall occur in the model.
Thus, it is this inward-outward
unbalance to increase Neptune's angular momentum.

\section{The Results and Concluding Remarks}

A semi-analytic model is employed to investigate the possible
outward migrations of Neptune. The driving mechanism for this
outward migration is the angular momentum exchange between
planetesimals and Neptune. Thus, the planetesimal sizes,
velocities, and inward-outward flux differences play important
roles for Neptune's migration histories. Different from the
previous work, we are able to link the exponential migration path
with the physical parameters of planetesimals in a clear picture.
The evolution of Neptune's orbital eccentricity is well addressed
and a final value close to the current observed one is obtained.

With the physically meaningful exponential paths of outward
migrations, the 3:2 and 2:1 resonant captures by Neptune are
studied. We find that for any exponential outward migrations, the
probabilities of resonant capture into 3:2 and 2:1 are the same.
When proto-KBOs are uniformly distributed in $R$ space initially,
we get results which are consistent with the results in Hahn and
Malhotra (2005): more proto-KBOs are captured into 2:1 than 3:2
resonance. However, in contrast, when proto-KBOs distribute as a
decay function of $R$, say $1/R^2$ or $1/R$, there are more
proto-KBOs captured into 3:2 than 2:1 resonances. This is more
consistent, though still different, with observations. According 
to data on the IAU Minor Planets Center 
(http://www.minorplanetcenter.net/iau/lists/TNOs.html), the 
number of KBOs trapped in the 3:2 resonance is over ten times 
more than the number in the 2:1 resonance.   The
differences between simulations and observations are huge, so the
problem remains unsolved. It is likely
that the current population ratio 
between 3:2 and 2:1 is mainly due to the original density distribution
of KBOs, which might be related to drag-induced inward 
migrations of proto-KBOs proposed in Jiang \& Yeh (2004, 2007).

\end{document}